\documentclass[12pt,twoside]{article}

\usepackage{epsf}
\usepackage{times}
\usepackage{a4wide}
\usepackage{epsfig}
\usepackage{fancyheadings,graphics}
\advance \headheight by 3.0truept       

\newlength{\nseparation}
\newenvironment{nfigure}[1]
        {\begin{figure}[#1]\hrule\vspace{\nseparation}\par}
        {\vspace{\nseparation}\par \hrule \end{figure}}

\usepackage{cite}   

\newcommand{\be}{\begin{equation}}
\newcommand{\ee}{\end{equation}}
\newcommand{\bea}{\begin{eqnarray}}
\newcommand{\eea}{\end{eqnarray}}
\newcommand{\bers}{\begin{eqnarray*}}
\newcommand{\eers}{\end{eqnarray*}}
\newcommand\un{\cal{U}}

\newcommand{\ov}[1]{\overline{#1}}
\newcommand{\eq}[1]{Eq.~(\ref{#1})}

\newcommand{\Bbar}{\,\overline{\!B}}

\newcommand{\bbs}{\ensuremath{B_s\!-\!\Bbar{}_s\,}}

\newcommand{\bra}[1]{\ensuremath{\langle #1 |}}
\newcommand{\ket}[1]{\ensuremath{| #1 \rangle }}

\begin{document}
\thispagestyle{empty}
\begin{flushright}
arXiv:0707.1535 (hep-ph)\\
July 2007
\end{flushright}
\vspace*{0.8cm}
\boldmath
\centerline{\LARGE\bf Unparticle physics effects in \bbs\ mixing}
\unboldmath
\vspace*{1.5cm}
\centerline{{\sc Alexander Lenz}}
\smallskip
\centerline{\sl Institut f{\"u}r Theoretische Physik -- 
                Universit{\"a}t Regensburg}
\centerline{\sl D-93040 Regensburg, Germany}
\vspace*{1cm}
\centerline{\bf Abstract}
\vspace*{0.3cm}
\noindent
We investigate unparticle effects in \bbs\ mixing. In particular we discuss
the possibility of reproducing the experimental result of $\Delta M_s$, while
having large effects on the mixing phase $\phi_s$, which might be visible
in current experiments.

\vspace*{1cm}


\vfill

\newpage
\setcounter{page}{1}
\pagenumbering{arabic}

\section{Introduction}
Unparticle physics has been recently suggested by Georgi 
\cite{georgi1,georgi2}.
Besides the standard model (SM) fields one assumes the existence of a 
non-trivial scale invariant sector at very high energies. 
These new fields with a nontrivial infrared fixed point are called 
Banks-Zaks fields (${\cal{BZ}}$) \cite{BZ}.
The SM fields and the $B{\cal Z}$ fields interact via the exchange of particles
of large mass $M_{\un}$. This interaction has the following generic form
\bers
\frac{1}{M_{\un}^k} O_{SM} O_{\cal{BZ}}\;, 
\eers 
where $O_{SM}$ is a operator of mass dimension $d_{SM}$  and 
 $O_{\cal{BZ}}$ is a operator of mass dimension $d_{\cal{BZ}}$
made out of SM  and ${\cal{BZ}}$ fields respectively.  
At a lower scale $\Lambda_{\un}$ the renormalizable
couplings of the ${\cal BZ}$ fields cause  dimensional
transmutation \cite{weinberg}. 
Below the scale $\Lambda_{\un}$ the ${\cal{BZ}}$ operators match
onto unparticle operators leading to new set interactions 
\bers
C_{\un}\frac{\Lambda_{\un}^{d_{\cal{BZ}}-d_{\un}}}{M_{\un}^k} O_{SM}
O_{\un}\;,
 \eers
where $C_{\un}$ is a coefficient of the low energy effective theory
and $O_{\un}$ is the unparticle operator with scaling dimension
$d_{\un}$. 
Georgi showed in \cite{georgi1} that unparticle stuff  with scale dimension 
$d_{\un}$ looks like a non-integral number $d_{\un}$ of invisible massless 
particles. 
Following the discussions in \cite{georgi1, georgi2} we only consider
two kinds of unparticles, scalar unparticles $O_{\un}$ and vector unparticles
$O_{\un}^{\mu}$. The coupling of
these unparticles to quarks is given as 
\be
\frac{c_S^{q'q}}{\Lambda_{\un}^{d_{\un}}}\bar q'
\gamma_\mu(1-\gamma_5) q~
\partial^\mu O_{\un}+\frac{c_V^{q'q}}{\Lambda_{\un}^{d_{\un}-1}}\bar q'
\gamma_\mu(1-\gamma_5) q~ O_{\un}^\mu+h.c.\;, \label{heff}
\ee 
where
 $c_{S,V}^{q'q}$ are flavor-dependent dimensionless coefficients.
We will consider the case $q=b$ and $q'=s$, which corresponds to
flavor changing neutral current (FCNC) transitions, giving contributions
to the \bbs\ mixing amplitude.
The propagator for the unparticle field is given as
\cite{georgi1, unprop} 
\begin{eqnarray} 
\int d^4 x e^{i P \cdot x}\langle 0 | TO_{\un}(x)
O_{\un}(0)|0 \rangle  
& = & 
i \frac{A_{d_{\un}}}{2 \sin d_{\un} \pi}
\frac{1} {(P^2+i \epsilon)^{2-d_{\un}}}e^{-i\phi_{\un}}\;, 
\\
\int d^4 x e^{i P \cdot x}\langle 0 | TO_{\un}^\mu(x) O_{\un}^\nu
(0)|0 \rangle 
& = & 
i \frac{A_{d_{\un}}}{2 \sin d_{\un} \pi}
\frac{-g^{\mu \nu} +P^\mu P^\nu/P^2} {(P^2+i
\epsilon)^{2-d_{\un}}}e^{-i \phi_{\un}}\;,
\end{eqnarray} 
with
\be A_{d_{\un}}= \frac{16 \pi^{5/2}}{(2 \pi)^{2 d_{\un}}}
\frac{\Gamma(d_{\un}+1/2)}{\Gamma(d_{\un}-1)\Gamma(2d_{\un})}\;,
 ~~~~{\rm and} ~~~\phi_{\un}=(d_{\un}-2)\pi \;.
\ee
Theoretical aspects of unparticles have been further discussed in 
\cite{unparticle}, while phenomenological consequences of unparticles have 
been investigated in \cite{unprop,unpheno,unmix1,unmix2,unmix3,unmix4}.
Mixing effects were already discussed in \cite{unmix1,unmix2,unmix3,unmix4},
where mostly effects on $\Delta M$ were studied. In this work we start 
from the fact that large new physics effects in $\Delta M$ are experimentally
excluded. Therefore we concentrate on new large contributions to the weak 
mixing phase.
In section 2.1 we introduce our notation and we review the status of the 
standard model predictions for the mixing quantities, in section 2.2
we discuss new physics effects to mixing in general. Section 2.3 contains our 
main results, the unparticle physics effects in \bbs\ mixing.
All our formulas are given for the $B_s$-system. The generalization to
the $B_d$-system is straightforward.
\section{B mixing}
\subsection{Notation and SM contributions to \bbs\ mixing}
\bbs\ oscillations are governed by a Schr\"odinger equation
\begin{equation}
i \frac{d}{dt}
\left(
\begin{array}{c}
\ket{B_s(t)} \\ \ket{\bar{B}_s (t)}
\end{array}
\right)
=
\left( M^s - \frac{i}{2} \Gamma^s \right)
\left(
\begin{array}{c}
\ket{B_s(t)} \\ \ket{\bar{B}_s (t)} 
\end{array}
\right)\label{sch}
\end{equation} 
with the mass matrix $M^s$ and the decay matrix $\Gamma^s$.  The
physical eigenstates $\ket{B_H}$ and $\ket{B_L}$ with the masses
$M_H,\,M_L$ and the decay rates $\Gamma_H,\,\Gamma_L$ are obtained by
diagonalizing $M^s-i \Gamma^s/2$.  The \bbs\ oscillations in
\eq{sch} involve the three physical quantities $|M_{12}^s|$,
$|\Gamma_{12}^s|$ and the CP phase 
$\phi_s=\arg(-M_{12}^s/\Gamma_{12}^s)$ (see e.g.\cite{run2}).  
$\Gamma_{12}$ stems from the absorptive part of the box diagrams - only
light internal particles like up and charm quarks contribute, 
while $M_{12}$ stems from
the dispersive part of the box diagram, therefore being sensitive to 
heavy internal  particles like the top quark or heavy new physics particles.
The calculable quantities {$  |M_{12}|$}, {$ |\Gamma_{12}|$} and 
{$ \phi = \mbox{arg}( -M_{12}/\Gamma_{12})$}
can be related to three observables (see \cite{LN2006,LN2006b,BBLN03} 
for a more detailed description):
\begin{itemize}
\vspace{-0.2cm}
\item \mbox{Mass difference 
       $ \Delta M = M^s_H -M^s_L \approx 2 { |M_{12}|} $}
\vspace{-0.2cm}
\item \mbox{Decay rate difference 
      $ \Delta \Gamma = \Gamma^s_L-\Gamma^s_H \approx 
        2 { |\Gamma_{12}| \cos  \phi } $}
\vspace{-0.2cm}
\item Flavor specific or semi-leptonic CP asymmetries:
      $a_{fs} =  { \mbox{Im} \frac{\Gamma_{12}}{M_{12}}} =
      \frac{\Delta \Gamma}{\Delta M} \tan \phi $.
\end{itemize}
Calculating the box diagram with internal top quarks one obtains
 \begin{equation}
        M_{12} =  \frac{G_F^2}{12 \pi^2} 
          (V_{ts}^* V_{tb})^2 M_W^2 S_0(x_t)
          {B_{B} f_{B_s}^2  M_{B_s}} \hat{\eta }_B
        \end{equation}
where $G_F$ is the Fermi constant, the $V_{ij}$'s are CKM elements, 
$ M_{B_s}$ and $M_W$ are the masses of $B_s$ meson and W boson.
The Inami-Lim function $S_0 (x_t = \bar{m}_t^2/M_W^2)$ 
\cite{IL} is the result of the box diagram 
without any gluon corrections. The NLO QCD correction is parameterized by 
$\hat{\eta}_B \approx 0.84$ \cite{BJW}.
The non-perturbative matrix element of the
four-quark operator
($\alpha,\beta=1,2,3$ are colour indices): 
\begin{eqnarray}
Q & =&   \ov s_\alpha \gamma_\mu (1-\gamma_5) b_\alpha \, 
         \ov s_\beta \gamma^\mu (1-\gamma_5) b_\beta . 
\label{defq} 
\end{eqnarray}
is parameterized by the 
bag parameter $B$ and the decay constant $f_{B_s}$.
\begin{eqnarray}
\bra{B_s} Q \ket{\ov B_s}  &=& \frac{8}{3} M^2_{B_s}\, f^2_{B_s} B 
    .  \label{defb} 
\end{eqnarray}
For our numerical estimates we will always use the input parameters listed
in \cite{LN2006}.
In \cite{LN2006} we obtained
\begin{equation}
\Delta M_s^{\rm Theo} \hspace{-0.2cm}= 19.3 \pm 6.4 \pm 1.9 \, \, \mbox{ps}^{-1}\, .
\end{equation}
The first error stems from the uncertainty in 
$f_{B_s}$ and the second error summarizes the remaining theoretical 
uncertainties.
This number has to be compared with the experimental value
\cite{deltamsexp}
\begin{equation}
\Delta M_s^{\rm Exp}  \hspace{-0.2cm}=  17.77 \pm 0.12  \, \, \mbox{ps}^{-1} \, .
\end{equation}
Due to our lack of a precise knowledge of $f_{B_s}$ there is still a 
sizeable room for new physics effects in $\Delta M_s$.

\begin{figure}[htb]
\begin{center}
\epsfig{file=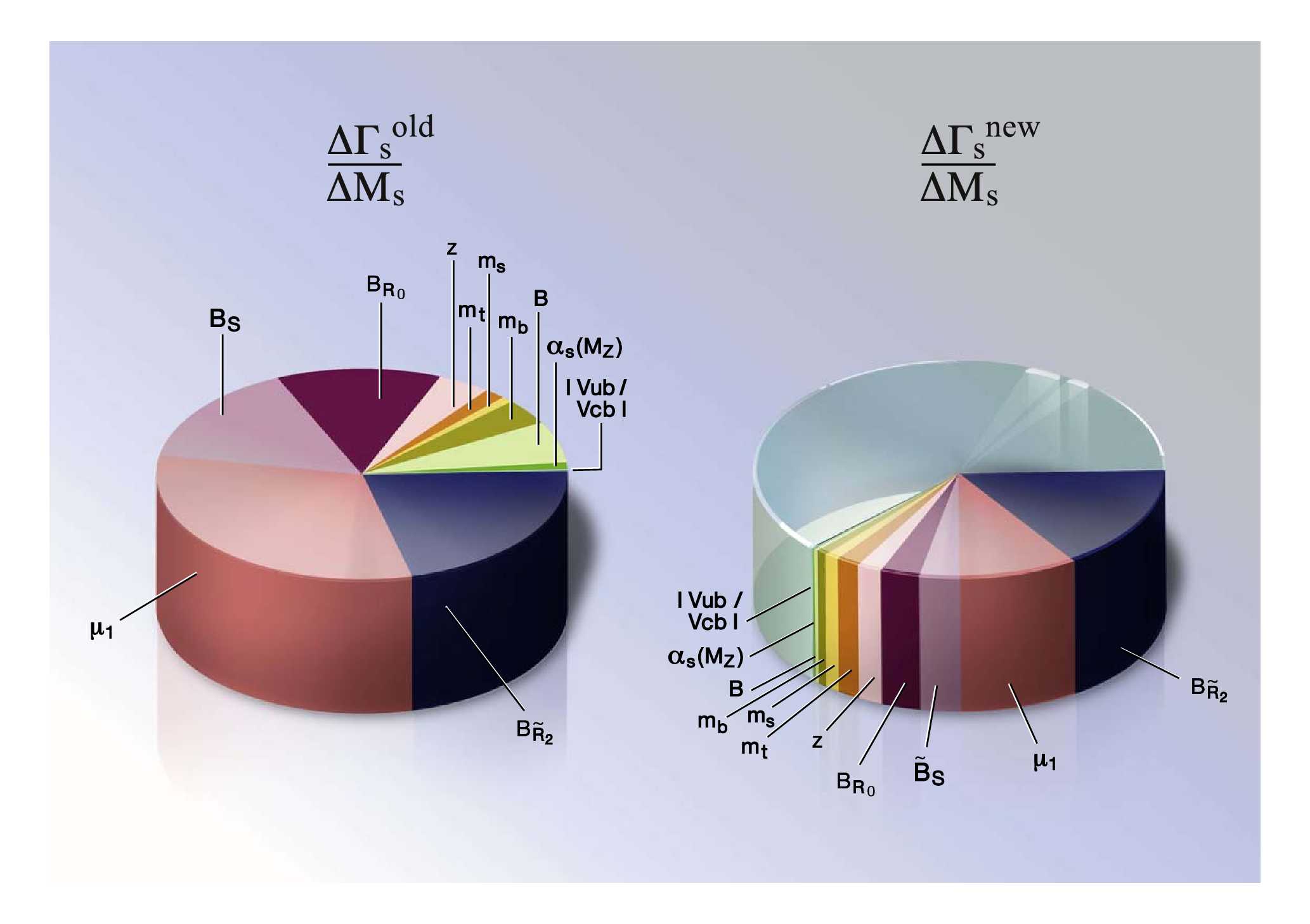,height=4.0 in}
\caption{Error budget for the theoretical determination of 
$\Delta \Gamma_s / \Delta M_s $. Compared to previous approaches (left)
the new strategy lead to a reduction of the theoretical error by almost 
a factor of 3.}
\label{fig:Kuchen}
\end{center}
\end{figure}

$\Gamma_{12}$ can be determined within the framework of the 
Heavy-Quark-Expansion (HQE) \cite{HQE}
as an expansion in $\Lambda/m_b$ and $\alpha_s$. The first contribution arises
at order $(\Lambda/m_b)^3$
\begin{equation}
\Gamma_{12} = 
\frac{\Lambda^3}{m_b^3} \left( \Gamma_3^{(0)} + \frac{\alpha_s}{4 \pi} \Gamma_3^{(1)} + \ldots\right) +
\frac{\Lambda^4}{m_b^4} \left( \Gamma_4^{(0)} + \ldots\right)
 +
\frac{\Lambda^5}{m_b^5} \left( \Gamma_5^{(0)} + \ldots\right)
+ \ldots  \, .
\end{equation}
The leading term $\Gamma_3^{(0)}$ was determined in \cite{dgLO}.
The numerical and conceptual important NLO-QCD corrections ($\Gamma_3^{(1)}$) 
were determined in \cite{BBGLN98,BBLN03}.
Subleading $1/m$-corrections, i.e. $\Gamma_4^{(0)}$ were calculated 
in \cite{BBD,1overm}
and even the Wilson coefficients of the $1/m^2$-corrections 
($\Gamma_5^{(0)}$) were calculated and found to be small \cite{LN2006}.
The smallness of these corrections was confirmed in \cite{Petrov2007}.
In addition to $Q$ now some new operators appear
\begin{eqnarray}
Q_S & = &  \ov{s}_\alpha (1+\gamma_5)  b_\alpha \, 
           \ov{s}_\beta  (1+\gamma_5)  b_\beta ,
\\
\widetilde{Q}_S & = &  \ov{s}_\alpha (1+\gamma_5)  b_\beta \, 
                       \ov{s}_\beta (1+\gamma_5)  b_\alpha \, .
\end{eqnarray}
We parameterise the matrix element of these operators as
\begin{eqnarray}
\bra{B_s} Q_S \ket{\Bbar_s} &=& -\frac{5}{3}  M^2_{B_s}\,
                  f^2_{B_s} B_S^\prime \, ,
\\
\bra{B_s} \widetilde Q_S \ket{\ov B_s} &=& \frac{1}{3}  M^2_{B_s}\,
                  f^2_{B_s} \widetilde B_S^\prime (\mu_2) \, , 
\end{eqnarray}
where we use the following abbreviation
\begin{eqnarray}
B_X^\prime &=& 
  \frac{M^2_{B_s}}{(\ov m_b+\ov m_s)^2} B_X \, .
\end{eqnarray}
In the vacuum insertion approximation (VIA) the bag factors $B$, $B_S$ and
$\widetilde B_S$ are equal to one. 
In \cite{LN2006} a strategy was worked out to reduce the theoretical 
uncertainty in $\Gamma_{12}/M_{12}$ by almost a factor of 3:
\begin{itemize}
\item The latest set of input parameters was used.
\item Logarithms of the type $z \ln z$ were summed up to all orders,
      c.f. \cite{BBGLN2002}.
\item Instead of the pole b-mass only the $\overline{MS}$-mass b-mass
      is used.
\item It was shown that the use of the operator basis $\{Q, \tilde{Q}_S \}$
      instead of $\{Q, {Q}_S \}$ leads to smaller theoretical uncertainties.
      While it is obvious that for $\Gamma_{12}/M_{12}$ the new basis has to
      be preferred - now the dominant part is completley free of 
      non-perturbative uncertainties, while in the old basis the dominat term
      was proportional to the ratio of the bag factors $B_S'$ and $B$ - it is
      not a priory clear what to prefer in the case of $\Gamma_{12}$. One 
      might think about averaging over the results in the two bases 
      \cite{cecilia}.
      Already this strategy reduces the error on $\Gamma_{12}$ 
      considerably.
      However having a closer look, one finds \cite{LN2006} that
      \begin{itemize}
      \item the numerical reduction of the $1/m_b$ corrections in the new
            basis is valid to all orders in $\alpha_s$
      \item the numerical correlations between $B$ and $B_S$ have not 
            been taken fully into account in the old basis.
      \end{itemize}
      Therefore we strongly suggest to use the new operator basis.
\end{itemize}
See Fig. (\ref{fig:Kuchen}) for an illustration of the improvements 
and \cite{old} for the shortcomings of the previous approach.
One gets
\begin{equation}
\frac{\Delta \Gamma_s}{\Delta M_s} = 
10^{-4} \cdot
\left[ 46.2  + 10.6 \frac{ B_S'}{B}  - 11.9  \frac{B_R}{B} 
\right] \, .
\end{equation}
where $B_R$ stands for the bag parameters of the $1/m_b$ parameters.
The dominant part of $\Delta \Gamma / \Delta M $ can now be determined without 
any hadronic uncertainties (for more details see \cite{LN2006})!
We obtained in \cite{LN2006} the following final numbers
with very conservative ranges for the input parameters
\begin{eqnarray}
\Delta \Gamma_s & \hspace{-0.6cm}= & \hspace{-0.6cm}
\left( 0.096   \pm 0.039 \right) \mbox{ps}^{-1},
\frac{\Delta \Gamma_s}{\Gamma_s}  
= 0.147 \pm 0.060,
\\
a_{fs}^s & \hspace{-0.6cm}= & \hspace{-0.6cm} 
\left( 2.06 \pm 0.57 \right) \cdot 10^{-5},
\frac{\Delta \Gamma_s}{\Delta M_s}  =  
\left( 49. 7 \pm 9.4 \right) 
 10^{-4} \! \! \! \! \! ,
\\
\phi_s &  \hspace{-0.1cm}= & 
0.0041 \pm 0.0008 \, \, \, = \, \, \, 0.24^\circ \pm 0.04^\circ \, .
\end{eqnarray}
The authors of \cite{Petrov2007} presented recently a number for
$\Delta \Gamma$ which is lower than the number above - 
but consistent within the errors.
Unfortunately the authors of \cite{Petrov2007} missed to include the
above mentioned theoretical improvements, therefore
their final number for $\Delta \Gamma_s$ has to be taken with a pinch of salt.
\subsection{General new physics contributions to \bbs\ mixing}
New physics (see e.g. \cite{NP}) is expected to have almost no impact 
on $\Gamma_{12}$ \cite{G1996}, (see \cite{Petrov2007,NP2} for some alternative 
viewpoints) but it can change $M_{12}$ considerably 
-- we denote the deviation factor by the 
complex number $\Delta$. Therefore one can write
\begin{eqnarray}
\Gamma_{12,s} & = & \Gamma_{12,s}^{\rm SM} \, ,
\\
M_{12,s}  & =  & M_{12,s}^{\rm SM} \cdot { \Delta_s} \, ;
\, \mbox{with} \,
{ \Delta_s} = { |\Delta_s|} e^{i  \phi^\Delta_s} \, .
\end{eqnarray}
\begin{nfigure}{tb}
\includegraphics[width=0.9\textwidth,angle=0]{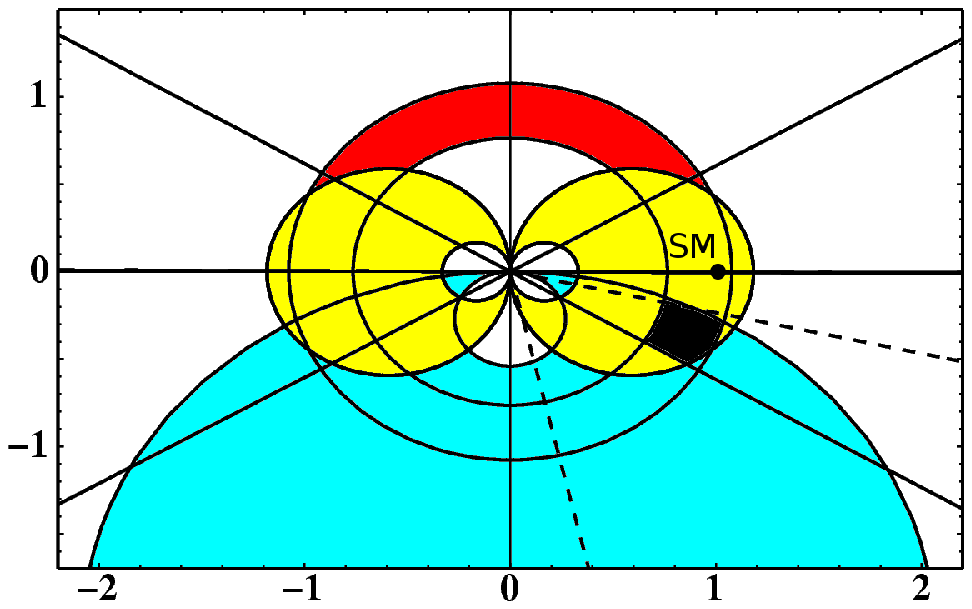}
\caption{Current experimental bounds in the complex $\Delta_s$-plane.
  The bound from $\Delta M_s$ is given by the red (dark-grey) ring around
  the origin. The bound from $\Delta \Gamma_s / \Delta M_s$ is given
  by the yellow (light-grey) region and the bound from $a_{fs}^s$ is given
  by the light-blue (grey) region. The angle $\phi_s^\Delta$ can be extracted
  from $\Delta \Gamma_s$ (solid lines) with a four fold ambiguity - one bound
  coincides with the x-axis! - or from the angular analysis in 
  $B_s \to J / \Psi \phi$ (dashed line). If the standard model is valid 
  all bounds should coincide in the point (1,0). The current experimental 
  situation shows a small deviation, which might become significant, if the 
  experimental uncertainties in $\Delta \Gamma_s$, $a_{\rm fs}^s$ and $\phi_s$ 
  will go down in near future.}\label{boundbandreal}
\end{nfigure}
With this parameterisation the physical mixing parameters can be written as
\begin{eqnarray}
 \Delta M_s  & =  & 2 | M_{12,s}^{\rm SM} | \cdot { |\Delta_s |} 
\label{bounddm},
\nonumber
\\
\Delta \Gamma_s   & =  &2 |\Gamma_{12,s}|
\cdot \cos \left( \phi_s^{\rm SM} + { \phi^\Delta_s} \right),
\label{bounddg}
\nonumber
\\
\frac{\Delta \Gamma_s}{\Delta M_s} 
& = &
 \frac{|\Gamma_{12,s}|}{|M_{12,s}^{\rm SM}|} 
\cdot \frac{\cos \left( \phi_s^{\rm SM} + { \phi^\Delta_s} \right)}
{ |\Delta_s|}
\label{bounddgdm},
\nonumber
\\
a_{fs}^s 
& = &
 \frac{|\Gamma_{12,s}|}{|M_{12,s}^{\rm SM}|} 
\cdot \frac{\sin \left( \phi_s^{\rm SM} + { \phi^\Delta_s} \right)}
{ |\Delta_s|}.
\label{boundafs}
\end{eqnarray}
Note that $\Gamma_{12,s} / M_{12,s}^{\rm SM}$ is now due to the above
mentioned improvements theoretically very well under control.
Next we combine the current experimental numbers with the theoretical 
predictions to extract bounds in the imaginary $\Delta_s$-plane by the use of
Eqs. (\ref{boundafs}), see Fig. (\ref{boundbandreal}). 
The width difference $\Delta \Gamma_s /\Gamma_s $ was 
investigated in \cite{dgexp}. 
The semi-leptonic CP asymmetry in the $B_s$ system has been determined 
in \cite{aslsexp} (see \cite{LN2006} for more details).
We use as experimental input the latest combination of the D0 collaboration 
\cite{combined}
\begin{eqnarray}
\Delta \Gamma_s & =  &0.17 \pm 0.09 \, \mbox{ps}^{-1} ,
\\
\phi_s  & = & -0.79  \pm 0.56 .
\\
a_{\rm fs}^{s} & = & \left(- 5.2 \pm 3.9 \right) \cdot 10^{-3} \, .
\end{eqnarray}
The HFAG \cite{HFAG} obtains the following combined value 
\begin{equation}
\Delta \Gamma_s = 0.071^{+0.053}_ {-0.057} \, \, \mbox{ps}^{-1} \, .
\end{equation}
The result of the comparison of experiment and theory 
can be seen in Fig. \ref{boundbandreal}, which is taken from \cite{LN2006}.
Already at this stage some hints for possible new physics contributions 
are visible, which manifest themselves as sizeable contributions to the mixing
phase $\phi_s$. In the next section we investigate, whether unparticle 
physics effects might create large contributions to $\phi_s$.
%
%
%
%
%
%
%
%
%
%
%
%
\subsection{Unparticle physics contributions to \bbs\ mixing}
In this section we determine possible contributions of unparticle physics effects
to \bbs\ mixing.
Using the operators from Eq. (\ref{heff}) we obtain for the
unparticle physics contribution to $M_{12}$
\begin{eqnarray}
M_{12}^{\cal U} = \frac{f_{B_s}^2}{M_{B_s}}
\frac{i e^{-i \phi_{\un}} A_{d_{\un}}}{4 \sin d_{\un} \pi}
\left\{
(c_V^{sb})^2 \left( \frac{M_{B_s}}{\Lambda_{\un}} \right)^{2 d_{\un} -2}
\left[ \frac{5}{3} \frac{m_b^2}{M_{B_s}^2} B_S' - \frac{8}{3} B \right]
+ 
(c_S^{sb})^2 \left( \frac{M_{B_s}}{\Lambda_{\un}} \right)^{2 d_{\un}}
\frac{5}{3} \frac{m_b^2}{M_{B_s}^2} B_S' 
\right\}
\label{m12unparticle}
\nonumber
\\
\end{eqnarray}
Using $\Delta M = 2 |M_{12}|$ and setting $m_b = M_{B_s}$ and $B = 1 =B_S'$ we 
reproduce the results in \cite{unmix2,unmix3,unmix4} for the mass differences.
With Eq. (\ref{m12unparticle}) we can determine $\Delta$:
\begin{equation}
\Delta = \frac{M_{12}}{M_{12}^{\rm SM}} = 1 + \frac{M_{12}^{\cal U}}{M_{12}^{\rm SM}}  
\end{equation}
Inserting the expressions for the unparticle propagator we obtain
\begin{eqnarray}
\frac{M_{12}^{\cal U}}{M_{12}^{\rm SM}} 
& = &
\frac{f_{B_s}^2 \pi^{\frac{1}{2}}}{M_{B_s} M_{12}^{\rm SM}}
\frac{(1+i \cot (d_{\un} \pi))  \Gamma(d_{\un} +\frac{1}{2})}
{\Gamma(d_{\un} -1 ) \Gamma(2 d_{\un} ) }
\left( \frac{M_{B_s}}{2 \pi \Lambda_{\un}} \right)^{2 d_{\un} -2}
\left\{
- (c_V^{sb})^2 
+ \frac{5}{3} (c_S^{sb})^2 \left( \frac{M_{B_s}}{\Lambda_{\un}} \right)^2 
\right\}
\nonumber \\
\end{eqnarray}
In order to simplify the expressions further we have to specify what
scale $\Lambda_{\un}$ we consider.
For simplicity we discuss only the cases  $\Lambda_{\un} = 1 $ TeV and  
$\Lambda_{\un} = 10 $ TeV.
\begin{eqnarray}
\Delta (\Lambda_{\un} = 1 \, {\rm TeV}) 
& \approx &
1 - \left[1+i \cot (d_{\un} \pi)\right] f_1(d_{\un})
\left\{ (c_V^{sb})^2  - 4.8 \cdot 10^{-5} (c_S^{sb})^2 \right\}
\\
\Delta (\Lambda_{\un} = 10 \, {\rm TeV}) 
& \approx &
1 - \left[1+i \cot (d_{\un} \pi)\right] f_{10} (d_{\un})
\left\{ (c_V^{sb})^2  - 4.8 \cdot 10^{-7} (c_S^{sb})^2 \right\}
\label{Deltaunparticle}
\end{eqnarray}
with
\begin{eqnarray}
f_1(d_{\un}) & = & 4.10 \cdot 10^{15} \frac{\Gamma(d_{\un} +\frac{1}{2})}
{\Gamma(d_{\un} -1 ) \Gamma(2 d_{\un} ) }
\left( 7.3 \cdot 10^{-7} \right)^{d_{\un}}
\\
f_{10} (d_{\un}) & = & 4.10 \cdot 10^{17} \frac{\Gamma(d_{\un} +\frac{1}{2})}
{\Gamma(d_{\un} -1 ) \Gamma(2 d_{\un} ) }
\left( 7.3 \cdot 10^{-9} \right)^{d_{\un}}
\end{eqnarray}
$f_1(d_{\un})$ is a strictly monotonic decreasing function, with e.g. 
$f_1(1.1) \approx 6 \cdot 10^7$ and $f_1(1.9) \approx 2 \cdot 10^3$.
$f_{10} (d_{\un})$ is also strictly monotonic decreasing, but yielding smaller 
values as $f_1$,  e.g. 
$f_{10}(1.1) \approx 4 \cdot 10^7$ and $f_{10}(1.9) \approx 35$.
If one assumes real couplings, then
the imaginary part of $\Delta$ is governed by the factor $\cot (d_{\un} \pi)$. This 
factor vanishes at $d_{\un} = 3/2$ and it is $\pm 1$ at $d_{\un} = 5/4, 7/4$.
Therefore one gets large imaginary contributions to $\Delta$ for values of  
$d_{\un} \in ]1;5/4[$ and $d_{\un} \in ]7/4;2[$.

For a further simplification we only consider a real coupling $c_V$ in the following.
The measurement of $\Delta M_s$ tells us that $|\Delta| = 0.92 \pm 0.32$.
Unparticle physics contributions yield
\begin{equation}
|\Delta | = \sqrt{1 - 2 c_v^2 f( d_{\un}) + [c_v^2 f( d_{\un})]^2 
[1 + \cot (d_{\un} \pi )]}
\end{equation}
The measurements of $\Delta \Gamma_s$, $\Delta \Gamma_s / \Delta M_s$, $a_{\rm fs}^s$
and $\phi_s$ give us information about $\phi_s$. Unparticle physics contributions 
yield 
\begin{equation}
\tan \phi_s = \frac{c_v^2 f( d_{\un}) \cot (d_{\un} \pi )}{1-c_v^2 f( d_{\un})}
\end{equation}
Demanding $|\Delta|$ to be equal to one is equal to adjusting the coupling $c_V$
in such a way that $c_v^2 f( d_{\un}) = 2/(1+ \cot (d_{\un} \pi ))$. We get then
\begin{equation}
\phi_s = \arctan \left[\frac{2}{1 - \tan (d_{\un} \pi )} \right]
\end{equation}
Plotting $\phi_s$ versus $d_{\un}$ one sees that large contributions to $\phi_s$
can be obtained, even if the constraint from $\Delta M_s$ is fulfilled exactly.
\begin{center}
\includegraphics{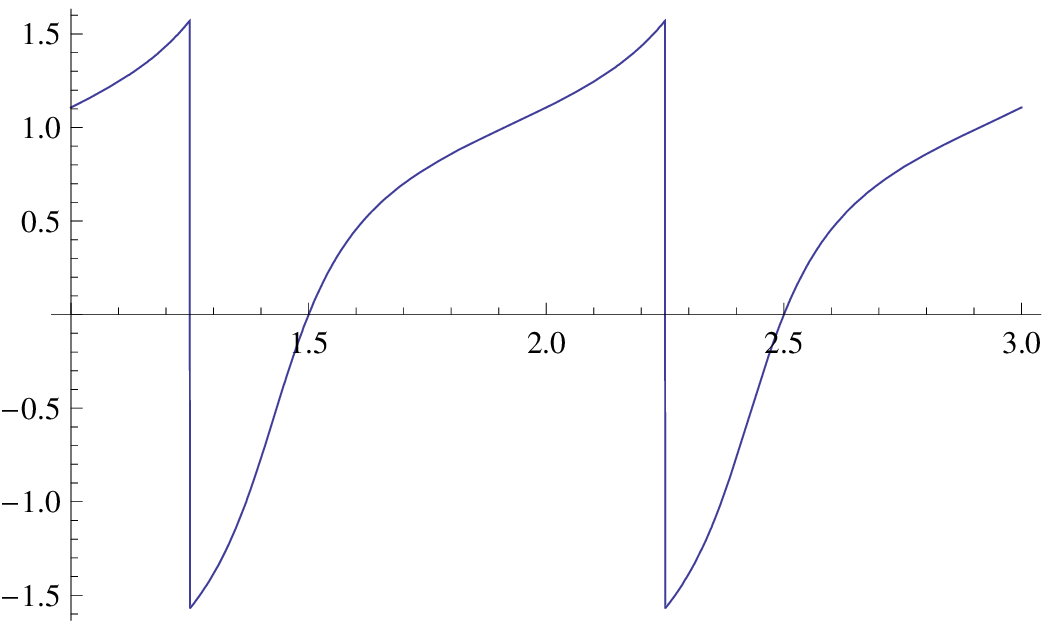}
\end{center}
\vspace{-6.5cm} \hspace{2cm}{\large $\phi_s$}
\begin{center}
\end{center}

\vspace{2.25cm} \hspace{12cm} {\large $d_{\un}$}
\begin{center}
\end{center}

\vspace{1.5cm}

From this figure one easily sees that large contributions to the mixing phase
$\phi_s$ can be created by choosing an appropriate scaling dimension $d_{\un}$:
E.g  $d_{\un} = 1.3975$ corresponds to $\phi^\Delta_s \approx - \pi / 4$ 
and $c_V = 7.3 \cdot 10^{-4}$ for 
$\Lambda_{\un} = 1 \, {\rm TeV} $ or $c_V = 1.8 \cdot 10^{-3}$ for 
$\Lambda_{\un} = 10 \, {\rm TeV} $.
This unparticle physics parameters yield $\Delta =1$ and $\phi_s = -\pi/4$ 
and therefore one would measure
\begin{eqnarray}
\Delta M_s = 17.4 \, \mbox{ps}^{-1},
&& \qquad
\Delta \Gamma_s = 0.068 \, \mbox{ps}^{-1},
\\
\frac{\Delta \Gamma_s}{\Delta M_s} = 3.91 \cdot 10^{-3},
&&\qquad
a_{\rm fs}^s = -3.89 \cdot 10^{-3} \, .
\end{eqnarray}
This corresponds to an enhancement of $-200$ in the case of $a_{\rm fs}$, while
$\Delta M_s$ stays close to the measured value.
Moreover if we assume  the following theoretical and experimental uncertainties: 
$\Delta M_s : \pm 15 \%$, 
$\Delta \phi_s : \pm 20 \%$, 
$\Delta \Gamma_s / \Delta M_s : \pm 15 \%$,
$a_{\rm fs}^s : \pm 20 \%$,
we obtain the regions in the $\Delta_s$-plane shown in figure \ref{boundband}.

\begin{nfigure}{tb}
\includegraphics[width=0.9\textwidth,angle=0]{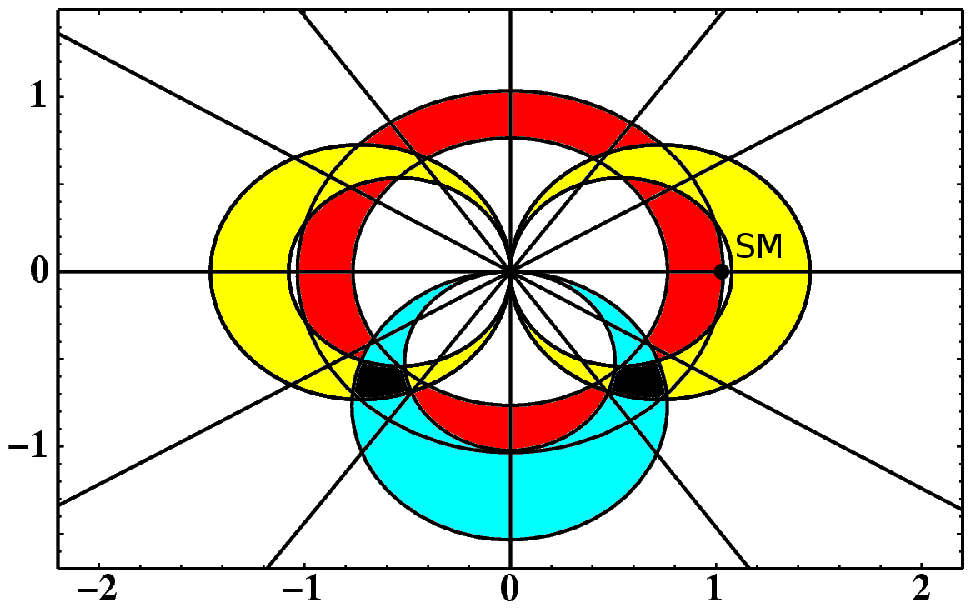}
\caption{Illustration of the bounds in the complex $\Delta_s$-plane 
  for $|\Delta_s| = 1$ and $\phi^\Delta_s = - \pi / 4$. 
  We assume the following overall uncertainties: 
  $\Delta M_s$                   (red or dark-grey)    : $\pm 15 \%$, 
  $\Delta \Gamma_s / \Delta M_s$ (yellow or light-grey): $\pm 15 \%$, 
  $a_{\rm fs}^s$                     (light-blue or grey)  : $\pm 20 \%$ and
  $\phi_s^\Delta$                (solid lines)         : $\pm 20 \%$.}
  \label{boundband}
\end{nfigure}

%
%
%
%
%
%

\section{Summary}\label{sect:sum}
In this paper we have investigated the unparticle effects to \bbs\ mixing.
We reproduce the results of \cite{unmix2,unmix3,unmix4} for the mass difference.
In contrast to these works, we concentrate on large effects of unparticle stuff on 
the weak mixing phase $\phi_s$, while contributions to $\Delta M_s$ are
small. The effects on the mixing phase vanish for
$d_{\un} = 3/2$ a case which was investigated in many previous unparticle physics 
analyses,
but it can be large for small deviations from 3/2. In particular we give an example
for a parameter set ($\Lambda_{\un}, d_{\un}, c_V$) for which one exactly reproduces
$\Delta M_s$, and one gets in addition large new physics effects in quantities
like the the semileptonic CP asymmetries. In the investigated case $a_{\rm fs}$ is 
enhanced by a factor of almost -200 compared to its SM value.
We are eagerly waiting for new experimental numbers to find out whether there is a 
sizeable mixing phase realized in nature.

\section*{Acknowledgements}
I would like to thank Miriam for providing me the necessary 
time to finish this project and J{\"u}rgen Rohrwild for proofreading the manuscript.
%
%
%
%
%
%
%
%
%


\begin{thebibliography}{99}
\bibitem{georgi1}
  H.~Georgi,
  Phys.\ Rev.\ Lett.\  {\bf 98} (2007) 221601
  [arXiv:hep-ph/0703260].

\bibitem{georgi2}
  H.~Georgi,
  arXiv:0704.2457 [hep-ph].

\bibitem{BZ}
  T.~Banks and A.~Zaks,
  Nucl.\ Phys.\  B {\bf 196} (1982) 189.

\bibitem{weinberg}
  S.~R.~Coleman and E.~Weinberg,
  Phys.\ Rev.\  D {\bf 7} (1973) 1888.

\bibitem{unprop}
  K.~Cheung, W.~Y.~Keung and T.~C.~Yuan,
  arXiv:0704.2588 [hep-ph];

\bibitem{unparticle}
  M.~A.~Stephanov,
  arXiv:0705.3049 [hep-ph];
  N.~V.~Krasnikov,
  arXiv:0707.1419 [hep-ph].
\bibitem{unpheno}
  G.~J.~Ding and M.~L.~Yan,
  arXiv:0705.0794 [hep-ph];
  Y.~Liao,
  arXiv:0705.0837 [hep-ph];
  T.~M.~Aliev, A.~S.~Cornell and N.~Gaur,
  arXiv:0705.1326 [hep-ph];
  M.~Duraisamy,
  arXiv:0705.2622 [hep-ph];
  C.~D.~Lu, W.~Wang and Y.~M.~Wang,
  arXiv:0705.2909 [hep-ph];
 P.~J.~Fox, A.~Rajaraman and Y.~Shirman,
  arXiv:0705.3092 [hep-ph];
  N.~Greiner,
  arXiv:0705.3518 [hep-ph];
  H.~Davoudiasl,
  arXiv:0705.3636 [hep-ph];
  D.~Choudhury, D.~K.~Ghosh and Mamta,
  arXiv:0705.3637 [hep-ph];
  S.~L.~Chen and X.~G.~He,
  arXiv:0705.3946 [hep-ph];
  T.~M.~Aliev, A.~S.~Cornell and N.~Gaur,
  arXiv:0705.4542 [hep-ph];
  P.~Mathews and V.~Ravindran,
  arXiv:0705.4599 [hep-ph];
  S.~Zhou,
  arXiv:0706.0302 [hep-ph];
  G.~J.~Ding and M.~L.~Yan,
  arXiv:0706.0325 [hep-ph];
  C.~H.~Chen and C.~Q.~Geng,
  arXiv:0706.0850 [hep-ph];
  Y.~Liao and J.~Y.~Liu,
  arXiv:0706.1284 [hep-ph];
  M.~Bander, J.~L.~Feng, A.~Rajaraman and Y.~Shirman,
  arXiv:0706.2677 [hep-ph];
  T.~G.~Rizzo,
  arXiv:0706.3025 [hep-ph];
  K.~Cheung, W.~Y.~Keung and T.~C.~Yuan,
  arXiv:0706.3155 [hep-ph];
  H.~Goldberg and P.~Nath,
  arXiv:0706.3898 [hep-ph];
  S.~L.~Chen, X.~G.~He and H.~C.~Tsai,
  arXiv:0707.0187 [hep-ph];
  R.~Zwicky,
  arXiv:0707.0677 [hep-ph];
  T.~Kikuchi and N.~Okada,
  arXiv:0707.0893 [hep-ph];
  C.~S.~Huang and X.~H.~Wu,
  arXiv:0707.1268 [hep-ph].

\bibitem{unmix1}
  M.~Luo and G.~Zhu,
  arXiv:0704.3532 [hep-ph].
\bibitem{unmix2}
  C.~H.~Chen and C.~Q.~Geng,
  arXiv:0705.0689 [hep-ph].
\bibitem{unmix3}
  X.~Q.~Li and Z.~T.~Wei,
  arXiv:0705.1821 [hep-ph].
\bibitem{unmix4}
  R.~Mohanta and A.~K.~Giri
  arXiv:0707.1234v1 [hep-ph]


\bibitem{run2}
K.~Anikeev {\it et al.},
\emph{$B$ physics at the Tevatron: Run II and beyond},
[hep-ph/0201071], Chapters 1.3 and 8.3.

\bibitem{LN2006}
  A.~Lenz and U.~Nierste,
  JHEP {\bf 06} (2007) 072
  [arXiv:hep-ph/0612167].

\bibitem{LN2006b}
  A.~Lenz,
{\it In the Proceedings of International Conference on Heavy Quarks and Leptons (HQL 06), Munich, Germany, 16-20 Oct 2006, pp 028}
  [arXiv:hep-ph/0612176];
  U.~Nierste,
  arXiv:hep-ph/0612310;
  A.~Lenz,
  arXiv:0705.3802 [hep-ph].

\bibitem{BBLN03}
  M.~Beneke, G.~Buchalla, A.~Lenz and U.~Nierste,
  Phys.\ Lett.\ B {\bf 576} (2003) 173
  [arXiv:hep-ph/0307344];
  M.~Ciuchini, E.~Franco, V.~Lubicz, F.~Mescia and C.~Tarantino,
  JHEP {\bf 0308} (2003) 031
  [arXiv:hep-ph/0308029].


\bibitem{IL}
  T.~Inami and C.~S.~Lim,
  Prog.\ Theor.\ Phys.\  {\bf 65} (1981) 297
  [Erratum-ibid.\  {\bf 65} (1981) 1772].


\bibitem{BJW}
  A.~J.~Buras, M.~Jamin and P.~H.~Weisz,
  Nucl.\ Phys.\ B {\bf 347} (1990) 491.


\bibitem{deltamsexp}
  A.~Abulencia {\it et al.}  [CDF Collaboration],
  arXiv:hep-ex/0609040;
A.~Abulencia  [CDF - Run II Collaboration],
 Phys.\ Rev.\ Lett.\  {\bf 97} (2006) 062003
  [arXiv:hep-ex/0606027];
 V.~M.~Abazov {\it et al.}  [D0 Collaboration],
  Phys.\ Rev.\ Lett.\  {\bf 97} (2006) 021802
  [arXiv:hep-ex/0603029].


\bibitem{HQE}
M.~A.~Shifman and M.~B.~Voloshin, in: \emph{Heavy Quarks}\ ed.\
V.~A.~Khoze and M.~A.~Shifman,
Sov.\ Phys.\ Usp.\  {\bf 26} (1983) 387;
M.~A.~Shifman and M.~B.~Voloshin,
Sov.\ J.\ Nucl.\ Phys.\  {\bf 41} (1985) 120
[Yad.\ Fiz.\  {\bf 41} (1985) 187];
M.~A.~Shifman and M.~B.~Voloshin,
Sov.\ Phys.\ JETP {\bf 64} (1986) 698
[Zh.\ Eksp.\ Teor.\ Fiz.\  {\bf 91} (1986) 1180];
J.~Chay, H.~Georgi and B.~Grinstein,
Phys.\ Lett.\ B {\bf 247} (1990) 399;
%
I.~I.~Bigi, N.~G.~Uraltsev and A.~I.~Vainshtein,
Phys.\ Lett.\ B {\bf 293} (1992) 430
[Erratum-ibid.\ B {\bf 297} (1992) 477];
  I.~I.~Y.~Bigi, M.~A.~Shifman, N.~G.~Uraltsev and A.~I.~Vainshtein,
  Phys.\ Rev.\ Lett.\  {\bf 71} (1993) 496
  [arXiv:hep-ph/9304225];
  B.~Blok, L.~Koyrakh, M.~A.~Shifman and A.~I.~Vainshtein,
  Phys.\ Rev.\ D {\bf 49} (1994) 3356
  [Erratum-ibid.\ D {\bf 50} (1994) 3572]
  [arXiv:hep-ph/9307247];
  A.~V.~Manohar and M.~B.~Wise,
  Phys.\ Rev.\ D {\bf 49} (1994) 1310
  [arXiv:hep-ph/9308246].


\bibitem{dgLO}
J.~S.~Hagelin and M.~B.~Wise,
Nucl.\ Phys.\ B {\bf 189} (1981) 87;
J.~S.~Hagelin,
Nucl.\ Phys.\ B {\bf 193} (1981) 123;
  E.~Franco, M.~Lusignoli and A.~Pugliese,
  Nucl.\ Phys.\  B {\bf 194} (1982) 403;
  L.~L.~Chau,
  Phys.\ Rept.\  {\bf 95} (1983) 1;
A.~J.~Buras, W.~Slominski and H.~Steger,
Nucl.\ Phys.\ B {\bf 245} (1984) 369;
  V.~A.~Khoze, M.~A.~Shifman, N.~G.~Uraltsev and M.~B.~Voloshin,
  Sov.\ J.\ Nucl.\ Phys.\  {\bf 46} (1987) 112
  [Yad.\ Fiz.\  {\bf 46} (1987) 181];
  A.~Datta, E.~A.~Paschos and U.~Turke,
  Phys.\ Lett.\  B {\bf 196} (1987) 382;
  A.~Datta, E.~A.~Paschos and Y.~L.~Wu,
  Nucl.\ Phys.\  B {\bf 311} (1988) 35.



\bibitem{BBGLN98}
  M.~Beneke, G.~Buchalla, C.~Greub, A.~Lenz and U.~Nierste,
  Phys.\ Lett.\ B {\bf 459} (1999) 631
  [arXiv:hep-ph/9808385].

\bibitem{BBD}
  M.~Beneke, G.~Buchalla and I.~Dunietz,
  Phys.\ Rev.\ D {\bf 54} (1996) 4419
  [arXiv:hep-ph/9605259].

\bibitem{1overm}
  A.~S.~Dighe, T.~Hurth, C.~S.~Kim and T.~Yoshikawa,
  Nucl.\ Phys.\ B {\bf 624} (2002) 377
  [arXiv:hep-ph/0109088].


\bibitem{Petrov2007}
  A.~Badin, F.~Gabbiani and A.~A.~Petrov,
  arXiv:0707.0294 [hep-ph].

\bibitem{BBGLN2002}
  M.~Beneke, G.~Buchalla, C.~Greub, A.~Lenz and U.~Nierste,
  Nucl.\ Phys.\  B {\bf 639} (2002) 389
  [arXiv:hep-ph/0202106].

\bibitem{cecilia}
  C.~Tarantino,
  arXiv:hep-ph/0702235.

\bibitem{old}
M.~Beneke and A.~Lenz,
J.\ Phys.\ G {\bf 27} (2001) 1219
[arXiv:hep-ph/0012222];
A.~Lenz,
arXiv:hep-ph/0412007;

\bibitem{NP}
  M.~Ciuchini and L.~Silvestrini,
  Phys.\ Rev.\ Lett.\  {\bf 97}, 021803 (2006)
  [arXiv:hep-ph/0603114];
  M.~Endo and S.~Mishima,
  Phys.\ Lett.\ B {\bf 640} (2006) 205
  [arXiv:hep-ph/0603251];
  Z.~Ligeti, M.~Papucci and G.~Perez,
  Phys.\ Rev.\ Lett.\  {\bf 97} (2006) 101801
  [arXiv:hep-ph/0604112];
  J.~Foster, K.~i.~Okumura and L.~Roszkowski,
  arXiv:hep-ph/0604121;
  P.~Ball and R.~Fleischer,
  arXiv:hep-ph/0604249;
  G.~Isidori and P.~Paradisi,
  Phys.\ Lett.\ B {\bf 639}, 499 (2006)
  [arXiv:hep-ph/0605012];
  S.~Khalil,
  Phys.\ Rev.\ D {\bf 74} (2006) 035005
  [arXiv:hep-ph/0605021];
  A.~Datta,
  Phys.\ Rev.\ D {\bf 74} (2006) 014022
  [arXiv:hep-ph/0605039];
  S.~Baek,
  JHEP {\bf 0609} (2006) 077
  [arXiv:hep-ph/0605182];
  X.~G.~He and G.~Valencia,
  Phys.\ Rev.\ D {\bf 74} (2006) 013011
  [arXiv:hep-ph/0605202];
  R.~Arnowitt, B.~Dutta, B.~Hu and S.~Oh,
  Phys.\ Lett.\ B {\bf 641} (2006) 305
  [arXiv:hep-ph/0606130];
  S.~Baek, J.~H.~Jeon and C.~S.~Kim,
  Phys.\ Lett.\ B {\bf 641} (2006) 183
  [arXiv:hep-ph/0607113];
  B.~Dutta and Y.~Mimura,
  arXiv:hep-ph/0607147;
  S.~Chang, C.~S.~Kim and J.~Song,
  arXiv:hep-ph/0607313;
  F.~J.~Botella, G.~C.~Branco and M.~Nebot,
  arXiv:hep-ph/0608100;
  S.~Nandi and J.~P.~Saha,
  arXiv:hep-ph/0608341;
  G.~Xiangdong, C.~S.~Li and L.~L.~Yang,
  arXiv:hep-ph/0609269;
  R.~M.~Wang, G.~R.~Lu, E.~K.~Wang and Y.~D.~Yang,
  arXiv:hep-ph/0609276;
  L.~x.~Lu and Z.~j.~Xiao,
  arXiv:hep-ph/0609279;
  M.~Blanke and A.~J.~Buras,
  arXiv:hep-ph/0610037;
  M.~Blanke, A.~J.~Buras, D.~Guadagnoli and C.~Tarantino,
  arXiv:hep-ph/0604057;
  M.~Blanke, A.~J.~Buras, A.~Poschenrieder, C.~Tarantino, S.~Uhlig and
  A.~Weiler, 
  arXiv:hep-ph/0605214;
  Z.~Ligeti, M.~Papucci and G.~Perez,
  Phys.\ Rev.\ Lett.\  {\bf 97} (2006) 101801
  [arXiv:hep-ph/0604112];
  Y.~Grossman, Y.~Nir and G.~Raz,
  Phys.\ Rev.\ Lett.\  {\bf 97} (2006) 151801
  [arXiv:hep-ph/0605028];
  S.~Jager and U.~Nierste,
  Eur.\ Phys.\ J.\  C {\bf 33} (2004) S256
  [arXiv:hep-ph/0312145].


\bibitem{G1996}
  Y.~Grossman,
  Phys.\ Lett.\ B {\bf 380} (1996) 99
  [arXiv:hep-ph/9603244].

\bibitem{NP2}
  E.~Lunghi and A.~Soni,
  arXiv:0707.0212 [hep-ph];
  S.~L.~Chen, X.~G.~He, A.~Hovhannisyan and H.~C.~Tsai,
  arXiv:0706.1100 [hep-ph];
  A.~Dighe, A.~Kundu and S.~Nandi,
  arXiv:0705.4547 [hep-ph].



\bibitem{dgexp}
R.~Barate {\it et al.}  [ALEPH Collaboration],
Phys.\ Lett.\ B {\bf 486} (2000) 286;
%
%
V.~M.~Abazov {\it et al.}  [D0 Collaboration],
arXiv:hep-ex/0702049;
%
D.~Acosta {\it et al.}  [CDF Collaboration],
%
Phys.\ Rev.\ Lett.\  {\bf 94} (2005) 101803
[arXiv:hep-ex/0412057];
%
A.~Abulencia  [CDF Collaboration],
arXiv:hep-ex/0607021;
%
V.~M.~Abazov {\it et al.}  [D0 Collaboration],
%
arXiv:hep-ex/0604046.
%
  V.~M.~Abazov {\it et al.}  [D0 Collaboration],
  Phys.\ Rev.\ Lett.\  {\bf 98} (2007) 121801
  [arXiv:hep-ex/0701012].




\bibitem{aslsexp}
  V.~M.~Abazov {\it et al.}  [D0 Collaboration],
  arXiv:hep-ex/0701007;
  V.~M.~Abazov {\it et al.}  [D0 Collaboration],
  Phys.\ Rev.\  D {\bf 74} (2006) 092001
  [arXiv:hep-ex/0609014].


\bibitem{combined}
  V.~Abazov  {\it et al.}  [D0 Collaboration],
  arXiv:hep-ex/0702030.


\bibitem{HFAG}
  E.~Barberio {\it et al.}  [Heavy Flavor Averaging Group (HFAG)
                  Collaboration],
  arXiv:0704.3575 [hep-ex].










\end{thebibliography}
\end{document}